\documentclass[aps,prfluids,reprint,onecolumn,groupedaddress]{revtex4-2}
\usepackage[pdftex]{graphicx}
\usepackage{lipsum}
\usepackage{SIunits}
\usepackage{amsmath}
\usepackage{xspace}
\usepackage{xcolor}

\newcommand{\etas}{\ensuremath{\eta_\mathrm{s}}\xspace}
\newcommand{\etar}{\ensuremath{\eta_\mathrm{r}}\xspace}
\newcommand{\phic}{\ensuremath{\phi_\mathrm{c}}\xspace}
\newcommand{\phistar}{\ensuremath{\phi^\star}\xspace}
\newcommand{\phicmono}{\ensuremath{\phi_{1}^\star}\xspace}
\newcommand{\phicbi}  {\ensuremath{\phi_{2}^\star}\xspace}
\newcommand{\phis}{\ensuremath{\phi_\mathrm{S}}\xspace}
\newcommand{\phil}{\ensuremath{\phi_\mathrm{L}}\xspace}
\newcommand{\ec}{\ensuremath{\dot\varepsilon_\mathrm{c}}\xspace}
\newcommand{\ecn}{\ensuremath{\dot\varepsilon_\mathrm{c,0}}\xspace}
\newcommand{\eloc}{\ensuremath{\dot\varepsilon_\mathrm{loc}}\xspace}
\newcommand{\edot}{\ensuremath{\dot\varepsilon}\xspace}
\newcommand{\ds}{\ensuremath{d_\mathrm{S}}\xspace}
\newcommand{\dl}{\ensuremath{d_\mathrm{L}}\xspace}
\newcommand{\al}{\ensuremath{a_\mathrm{L}}\xspace}
\newcommand{\as}{\ensuremath{a_\mathrm{S}}\xspace}
\newcommand{\di}{\ensuremath{d_\mathrm{I}}\xspace}
\newcommand{\hd}{\ensuremath{\hat d}\xspace}
\newcommand{\hds}{\ensuremath{\hat{\ds}}\xspace}
\newcommand{\hdl}{\ensuremath{\hat{\dl}}\xspace}
\newcommand{\hdi}{\ensuremath{\hat{\di}}\xspace}
\newcommand{\hdelta}{\ensuremath{\hat\delta}\xspace}
\newcommand{\hdeltai}{\ensuremath{\hat\delta_\mathrm{I}}\xspace}

\begin{document}

%\preprint{}

\title{Caging and fluid deformations in dense bidisperse suspensions}


\author{Virgile Thi\'evenaz}
\affiliation{PMMH, CNRS, ESPCI Paris, Universit\'e PSL, Sorbonne Universit\'e, Universit\'e Paris Cit\'e, F-75005, Paris, France}
% \affiliation{Department of Mechanical Engineering, University of California, Santa Barbara, California 93106, USA}
\author{Nathan Vani}
\affiliation{PMMH, CNRS, ESPCI Paris, Universit\'e PSL, Sorbonne Universit\'e, Universit\'e Paris Cit\'e, F-75005, Paris, France}
% \affiliation{Department of Mechanical Engineering, University of California, Santa Barbara, California 93106, USA}
\author{Alban Sauret}
\affiliation{Department of Mechanical Engineering, University of California, Santa Barbara, California 93106, USA}
\affiliation{Department of Mechanical Engineering, University of Maryland, College Park, Maryland 20742, United States}
%\email[]{}
%\thanks{}


\date{\today}

\begin{abstract}
    We investigate the link between the geometric environment of particles,
    the local deformations of the solvent, and the bulk effective viscosity
    in non-Brownian suspensions.
    First, we discuss the caging of particles by their neighbors,
    and especially the caging of small particles by large ones in bidisperse suspensions.
    We develop a model that attributes an effective volume to particles 
    depending on their environment,
    and yields the local deformations and effective viscosity.
    We compare this model to data from the literature, as well as to our own
    experiments with suspensions of non-Brownian polystyrene beads.
    Using dissolved polymers and their coil-stretch transition as strain probes,
    we measure the local deformation of the liquid and the effect of caging thereon.
    We obtain a linear relationship between the amplified local strain rate and 
    the particle volume fraction, in which the critical volume fraction \phic 
    appears as an effective volume of the particles;
    this relationship is found valid up into the dense regime.
\end{abstract}

%\keywords{}

\maketitle

% Introduction
Many fluids in nature and industry are suspensions
of solid particles in a liquid~\cite{kostynick2022rheology},
and as such are a challenge to describe in terms of continuous media~\cite{ness2022physics}.
Rigid particles deform the flow and increase the stress applied on
the liquid~\cite{einstein1905}. 
As a result, the suspension exhibits an effective viscosity \etas 
which may be orders of magnitude higher than that of the liquid $\eta_0$~\cite{guazzelli2018}.
Many semi-empirical laws describe how the relative viscosity $\etar = \etas/\eta_0$ 
increases with the volume fraction of solid $\phi$
\cite{eilers1941viskositat, krieger1959, frankel1967viscosity};
for example that of Maron and Pierce~\cite{maron1956}: $\etar = (1-\phi/\phic)^{-2}$.
All introduce an empirical parameter \phic, 
described as the critical volume fraction at which the viscosity diverges.
It is remarkable that these laws describe well many different suspensions
-- particles of different sizes and materials~\cite{guazzelli2018}, 
of different shapes~\cite{bounoua2019shear},
in different solvents~\cite{nguyen-le2023} --
all the differences being contained in this parameter \phic.

Polydispersity plays a crucial role, and most common suspensions are polydisperse,
meaning their grains have a range of size. 
Such suspensions are less viscous than ideal monodisperse suspensions of equal volume 
fraction~\cite{shapiro1992,pednekar2018}.
Recent studies on the flow of bidisperse suspensions (containing particles of two different sizes)
have shown that a minute amount of particles of a second size 
can change the dynamics significantly~\cite{thievenaz2021b, jeong2022dip, pelosse2023probing},
by modifying the scale of collective motion~\cite{thievenaz2022}.
Shear-induced migration~\cite{lyon1998experimental1} causes segregation when a bidisperse
suspension flows confined in a channel~\cite{lyon1998experimental2, nath2022flow}.
The Wyart-Cates model~\cite{wyart2014},
which explains shear-thickening as a transition
from frictionless to frictional contacts, is successful at describing monodisperse suspensions 
but fails with bidisperse suspensions~\cite{guy2020testing}.
Even bidisperse active matter shows peculiar behavior because of 
differentiated interactions between small and large particles~\cite{maity2023spontaneous}.
Despite this apparent complexity, the same empirical laws match the viscosity of
bidisperse suspensions and simply yield other values of \phic~\cite{gondret1997}.
How to estimate \phic, and how it relates to the jamming transition,
remain open questions.

In this Letter, we investigate the local environment of particles in bidisperse 
non-Brownian suspensions, especially the caging of small particles by large ones.
We infer a model for the local strain rate.
Then, using the coil-stretch transition of dissolved polymers as a strain probe
\cite{thievenaz2021a,rajesh2022transition}, 
we measure the local strain rate in the liquid phase.
Finally, we discuss the effect of caging on the macroscopic viscosity.
For the sake of clarity, we begin by describing the model before our experiments
and their discussion.

%%%%%%%%%%%%%%%%%%%%%%%%%%%%%%%%%%%%%%%%%%%%%%%%%%%%%%%%
% The model
\begin{figure}[t]
    \centering
    \includegraphics[width=\linewidth]{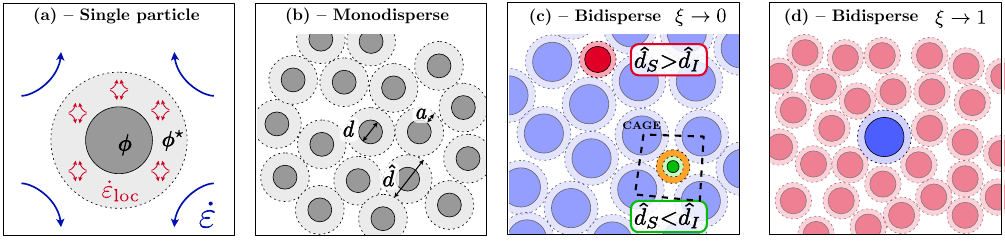}
    \caption{(a) Amplification of the macroscopic strain rate \edot near the particle; 
    within the effective volume \phistar, the strain rate in increased up to \eloc.
    (b) Monodisperse suspension of spheres (dark gray) and their effective volume (light gray).
    (c) Bidisperse suspension with many large particles and few small ones ($\xi \to 0$).
    Depending on the relative size of the small particles \ds and of the interstices between
    the large particles \di, the small particles may (in green) or may not (in red) be caged by
    the large ones. In the former case, the small particles have an effective volume equal to that
    of the interstice (in orange).
    (d) Bidisperse suspension with few large particles and many small ones ($\xi \to 1$).
    }
    \label{fig:model}
\end{figure}

The model consists in attributing an effective volume to each particle, 
so that the suspension can be seen as an effective packing of spherical cells.
Each cell contains a particle and some surrounding fluid;
the solid volume fraction is $\phi$, the fraction of effective volume is \phistar.
Particles are rigid, only the liquid may deform.
Therefore, when the suspension deforms at a macroscopic strain rate \edot,
the liquid deforms at an increased local strain rate \eloc.
Figure~\ref{fig:model}a represents an idealized vision of this mechanism;
in reality the size and shape of the region of amplified strain depends on the particle
environment and on the microstructure, which depends on the particle size distribution,
on the flow boudary conditions, etc.
We consider that the strain rate can be averaged over the effective volume,
so the total deformation is
$\edot\phistar = \eloc(\phistar-\phi)$,
hence
\begin{equation}
    \eloc(\phi) / \edot = \left( 1 - \phi/\phistar \right)^{-1}.
    \label{eq:local}
\end{equation}
Eq.~(\ref{eq:local}) was obtained by Mills and Snabre following similar arguments
\cite{mills1985non,mills1988} and served as an hypothesis to build more complex rheological
models.
We shall demonstrate that this equation is directly validated by experiments
by measuring both \edot and \eloc.
But first, we discuss how the effective volume vary in bidisperse suspensions.

% Bidisperse cell packing
A bidisperse suspension contains particles of two different sizes;
large ones of diameter \dl and volume fraction \phil,
and small ones of diameter \ds and volume fraction \phis;
$\phi = \phis + \phil$.
We introduce the size ratio of the particles $\delta = \dl/\ds$,
the size ratio of the cells $\hdelta = \hdl/\hds$,
and the share of small particles $\xi = \phis/\phi$.
In a monodisperse suspension, all cells have on average the same size (Fig.~\ref{fig:model}b);
each cell contains one particle of diameter $d$ and the layer of liquid of thickness $a$,
so the cell diameter is simply $\hd = d+2a$.
In a bidisperse suspension, $a$ may depend on the size and 
local environment of the particle considered, 
so $\hd_i = d_i + 2a_i$ where $i$ refers to a given particle.
The average effective volume of the bidisperse suspension is then
the compacity of the packing of the bidisperse cells \phicbi.
We compute it using the model of Ouchiyama and Tanaka~\cite{ouchiyama1981,ouchiyama1984},
which consists in evaluating the local compacity around a given sphere surrounded 
with spheres of the average size, and then taking the average over the size distribution.
This model has been adapted to bidisperse packings~\cite{gondret1997,thievenaz2021b},
and agrees well with measurements for dry grains~\cite{kalyon2014factors}.
Thanks to it, \phicbi can be related to the compacity of a monodisperse packing \phicmono:
\begin{equation}
    \phicbi = f(\phicmono, \hdelta, \xi).
    \label{eq:phicbi}
\end{equation}
Since the full expression of \phicbi is quite long and not necessary for the discussion,
we leave it to the supplementary material~\cite{supmat}.

\begin{figure*}[t!]
    \centering
    \includegraphics[width=.99\linewidth]{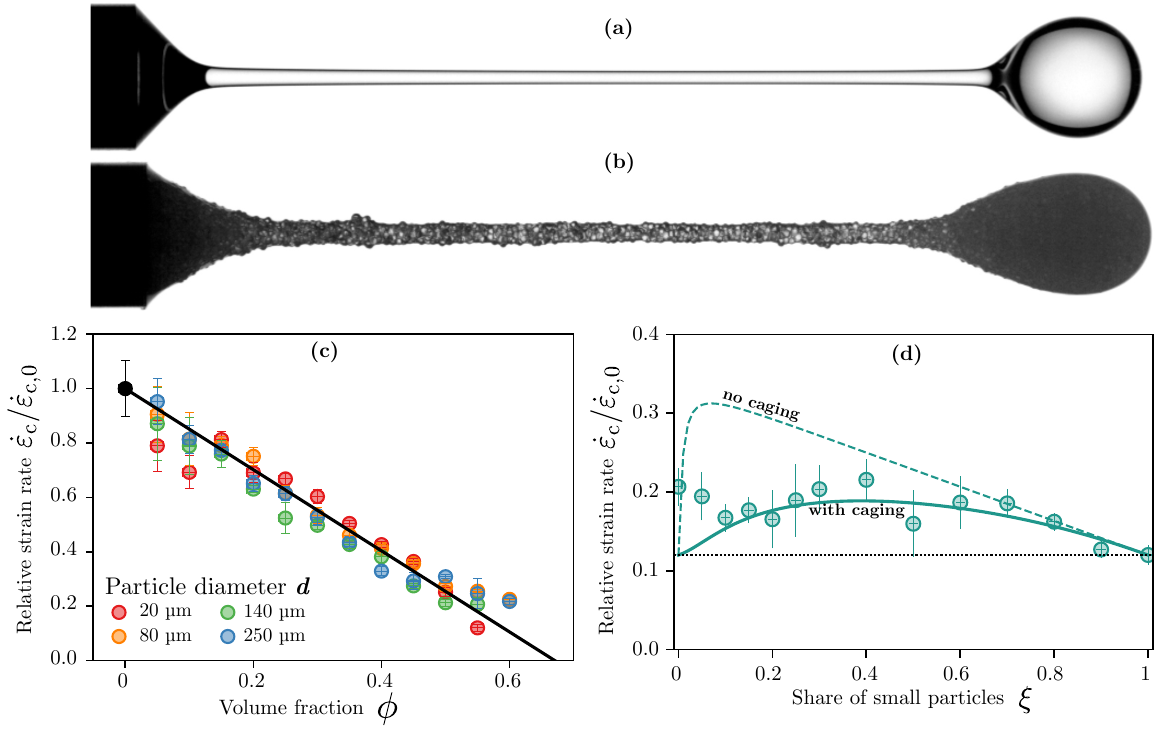}
    \caption{(a) Stretching of a polymer solution.
        The nozzle is \unit{5.5}\milli\meter-wide.
        (b) Same experiment with \unit{250}\micro\meter-polystyrene particles at $\phi=60\%$.
        (c) Amplification of the local rate of strain $\ec/\ecn$\ \textit{vs.}
        volume fraction for monodisperse suspensions.
        (d)$\ec/\ecn$ \emph{vs.}
        the share of small particles $\xi$ in bidisperse suspensions,
        with $\ds=\unit{20}\micro\meter$ and $\dl=\unit{140}\micro\meter$, 
        at constant $\phi=55\%$.
        Lines represent the model including the caging effect (solid line) and
        excluding it (dashed line).
    }
    \label{fig:strain}
\end{figure*}

% Computing δ
Eq.~(\ref{eq:phicbi}) requires that we know \hdelta.
There are two different cases, depending on whether the small particles are large enough
to disturb the packing of large cells, or are small enough to sit in the interstices
between the large cells.
% δ = 1
For particles of similar size ($\delta \simeq 1$), 
we assume that $a_i$ is simply proportional to $d_i$;
$\al/\dl = \as/\ds$.
The size ratio of cells is then equal to the size ratio of particles: $\hdelta = \delta$.
Physically, this means that the small particles experience the same environment as
the large ones. 
 
% δ > 1
If the small particles are small enough, their cells may sit in the interstices between 
large cells.
Such interstices should be surrounded by six large cells on average, two per dimension of space.
Assuming the interstice has no privileged direction, it should be an octaedric site, 
of diameter $\hdi = \left(\sqrt2-1\right)\hdl$.
% This is only valid at low strain rates; were the strain rate too high there would be 
% a privileged direction and symmetry would be broken.
The size ratio of the large cell to the interstitial cell is then :
\begin{equation}
    \hdeltai = \hdl / \hdi = \left(\sqrt{2}-1\right)^{-1} \simeq 2.41.
    \label{eq:di}
\end{equation}
We model the interstices as follows.
There are two possible environments for a small particle:
\textit{free} among its neighbors, or \textit{caged} by them
(Fig.~\ref{fig:model}(c), respectively in red and green).
We assume that if a small particle is caged in an interstice 
then its effective volume is the whole interstice 
(Fig.~\ref{fig:model}(c), green particle in orange interstice).
Therefore, its effective diameter is \hds if it is free and \hdi if it is caged.
Only small particles can be caged, but if there are more small particles
than cages, then some small particles will be free.
Assuming there are as many interstices as large cells, 
the average effective diameter of the small particles is:
\begin{equation}
    \hds(\xi) = \xi(\hdl/\delta)+ (1-\xi)\hdi.
    \label{eq:hds}
\end{equation}
Therefore, the general expression of the size ratio of cells is:
\begin{align}
    \hdelta &= \delta \quad \text{ for } \quad \delta\le\hdeltai;
    \label{eq:delta1}
    \\
    \hdelta &= \frac{\delta \hdeltai}{\xi\hdeltai + (1-\xi)\delta} 
    \quad \text{ for }\quad  
    \delta>\hdeltai.
    \label{eq:delta2}
\end{align}
In the limit $\xi \to 1$, both equations are equivalent;
it means that if there are few large particles, caging the small ones is impossible
(Fig.~\ref{fig:model}d).
Eqs.~(\ref{eq:delta1}-\ref{eq:delta2}) together with Eq.~(\ref{eq:phicbi}) enable to compute
the effect of caging on the effective volume in bidisperse suspensions.
The local strain rate and the bulk viscosity can thence be deduced.

%%%%%%%%%%%%%%%%%%%%%%%%%%%%%%%%%%%%%%%%%%%%%%%%
% Strain rate amplification

% Measuring the amplification of the strain rate
Our original measure of the local strain rate \eloc consists in adding minute amounts of polymer
dissolved in the liquid phase and then stretching the suspension.
A most simple stretching experiment consists in observing the detachment of a drop
(Figure~\ref{fig:strain}a and b).
At rest, polymer chains are coiled since this conformation maximizes their entropy;
above a critical rate of strain the chains unwind and stretch;
this is the coil-stretch transition~\cite{de-gennes1974,smith1999single}.
Macroscopically, the thinning of a polymer solution switches from 
a Newtonian to a viscoelastic regime~\cite{amarouchene2001},
through a universal transition that is only controlled by the critical macroscopic 
strain rate \ec~\cite{rajesh2022transition}.
With particles (large compared to the polymer chains) in the solution,
the transition occurs for a lower \ec,
meaning that the polymer chains undergo higher stress~\cite{thievenaz2021a}. 
Therefore, by comparing the critical strain rate for the suspension \ec and
for the polymer solution without particles \ecn, 
we can measure by how much particles amplify the local strain.
Since coil-stretch transition marks the onset of significant viscoelastic behavior,
the microstructure before and at the transition should not be affected by viscoelasticity.

\begin{figure*}[t!]
    \centering
    \includegraphics[width=\linewidth]{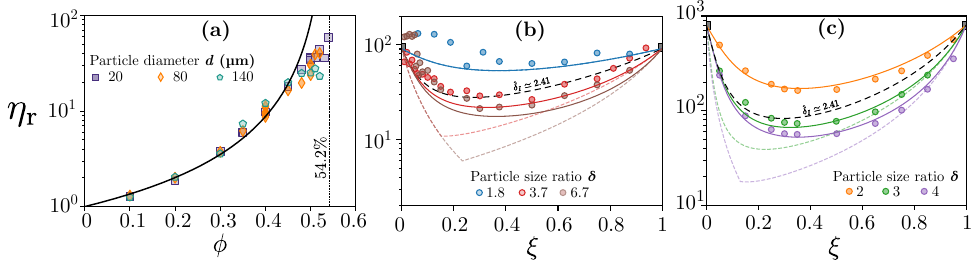}
    \caption{
        (a) Relative viscosity \etar of monodisperse suspensions \emph{vs.} volume fraction $\phi$;
        the line indicates the best fit by Eq.~(\ref{eq:visc}).
        (b) \etar \emph{vs.} the share of small particles $\xi$ for bidisperse suspensions 
        at $\phi=50\%$.
        (c) Comparisons with the simulations of Pednekar~\emph{et al.}~\cite{pednekar2018}
        for $\phi=60\%$.
        For both (b) and (c), solid lines represent the predictions of the model.
        Colored dashed lines represent the prediction not taking caging into account.
        The black dashed lines represent the case $\hdelta=\hdeltai$,
        \textit{i.e.} the asymptotical regime $\xi \to 0$.
        }
    \label{fig:visc}
\end{figure*}

% The experiments
The polymer solution is polyethylene oxide (PEO, Sigma)
of molar weight \unit{4\cdot 10^6}\mole\per\kilo\gram,
dissolved at a weight fraction of 0.02\% in a 75\%-water, 25\%-glycerol mixture.
It is gently stirred on a roller mixer during 
two days to ensure uniform concentration, and then mixed with the particles.
We used polystyrene beads (Dynoseeds from Microbeads) 
of four different diameters (20, 80, 140 and \unit{250}\micro\meter)~\cite{supmat}.
These beads are rather smooth, with a roughness of order \unit{100}\nano\meter~\cite{deboeuf2009particle};
in aqueous solvents they have repulsive interactions~\cite{nguyen-le2023}.
For the stretching experiments, 
the suspensions were extruded through a \unit{5.5}\milli\meter-wide nozzle.
Based on our previous work~\cite{thievenaz2021a, rajesh2022transition}, 
this system satisfies the following constraints:
neutrally-buoyant particles;
sharp transition to the viscoelastic regime to maximize the precision on \ec;
macroscopic transition to the viscoelastic regime at flow scales much larger
than the particle size, to avoid confinement~\cite{supmat}.
The thinning dynamics are recorded using a high-speed camera (Phantom VEO 710) 
and a macro lens (Nikon Micro-Nikkor AI-s 200mm f/4).
The macroscopic strain rate $\edot(t)$ is computed from the thickness $h(t)$ of the liquid thread
at its thinnest point; $\edot(t) = - 2\:\mathrm{d}(\log h)/ \mathrm{d} t$~\cite{supmat}.
The critical strain rate \ec is the maximum value of $\dot\varepsilon$.
Each experiment was repeated ten times to reduce uncertainty.

Using this protocol, we measured \ec for monodisperse suspensions 
with volume fraction $\phi$ ranging from 5\% to 60\%,
and for the solution without particles: $\ecn = \ec(\phi\!=\!0)$.
Since the coil-stretch transition is caused by the flow at the scale of the polymer chains,
which is much smaller than the particles, the local strain rate \eloc should have 
the same value in all experiments; therefore $\eloc=\ecn$.
Figure~\ref{fig:strain}(c) shows that the ratio $\ec/\ecn$
follows a linear decrease: $\ec/\ecn = 1 - \phi/\phistar$; the best fit gives $\phistar$=67\%.
We observe no effect of the particle size.
Since $\ec/\ecn$ is the ratio of macroscopic to local strain rate,
this is a direct validation of Eq.~(\ref{eq:local}), which is therefore no longer an 
hypothesis but an experimental fact.
It must be emphasized that Eq.~(\ref{eq:local}) is valid throughout the whole 
range of volume fractions, from 5\% up to 60\%.

This method can be extended to bidisperse suspensions by mixing beads of two sizes 
in the same polymer solution.
We conducted many experiments with various particle size couples and 
different volume fractions,
but because the effect of the size distribution on \ec is weak we could only observe it
in the most concentrated case ($\phi=55\%$) with the highest size ratio
($\delta=6.7$, see Fig.~\ref{fig:strain}(d)).
It should be noted that for such a high size ratio caging should be important.
The full line represents the prediction of Eqs.~(\ref{eq:phicbi}-\ref{eq:delta2}),
in very good agreement with the data.
The value of \phicmono is taken to match the value of the point at $\xi=1$.
The dashed line represents the prediction not taking caging into account,
that is taking cell sizes proportional to particle sizes;
it is off by a factor 2.
Caging is therefore important.
The less good agreement when $\xi\to 0$ is likely due to our hypothesis that
\phicmono is exactly the same for the different sizes;
the difference is small yet not completely negligible.

%%%%%%%%%%%%%%%%%%%%%%%%%%%%%%%%%%%%%%%%%%%%%%%%
% Viscosity
Although the experimental measurement of the strain rate is quite useful,
the usually measured quantity is the viscosity of the suspension \etas.
It may be derived by writing that the power dissipated per unit volume
of suspension equals the power dissipated per unit volume of liquid 
multiplied by the volume fraction of liquid:
\begin{equation}
    \etas\edot^2 = (1-\phi)\eta_0\eloc^2,
    \label{eq:conserv}
\end{equation}
hence the relative viscosity~\cite{mills1985non}:
\begin{equation}
    \etar = \frac{\etas}{\eta_0} = \frac{1-\phi}{\left(1 - \phi/\phistar\right)^2}.
    \label{eq:visc}
\end{equation}

We measured the shear viscosity of suspensions of the same polystyrene beads,
dispersed in a mixture of water and polyethylene glycol (58\%-42\% wt.).
This liquid has the same density as the beads (\unit{1050}\kilogram\per\cubic\meter);
it is Newtonian at the shear rate considered and its viscosity is 
\unit{103}\milli\pascal\cdot\second.
The rheometer (Anton Paar MCR92) is equipped with a plane-plane geometry with rough surfaces
to avoid slippage; the gap between the plates is \unit{1}\milli\meter.
The viscosity is measured in the stationary regime under an imposed shear rate of 
\unit{100}\second\reciprocal.
For monodisperse suspensions, Eq.~(\ref{eq:visc}) matches our measurements up to $\phi \simeq 48\%$;
the best fit in this range yields $\phistar=54.2\%$ (Fig.~\ref{fig:visc}a),
the same value for the three particle sizes.
For bidisperse suspensions, we compare our model to viscosity measurements for $\phi=50\%$
(Fig.~\ref{fig:visc}b); we obtain a fair agreement with our experiments,
taking $\phicmono=54.2\%$ as measured on monodisperse suspensions.
The most remarkable result comes from the comparison with the numerical simulations 
of Pednekar~\emph{et al.}~\cite{pednekar2018} (Fig.~\ref{fig:visc}c),
which describe the most concentrated case of $\phi=60\%$.
For $\delta=2$ (orange), the agreement is perfect;
it should be noted that in this case there is no caging, since $\delta<\hdeltai$.
We obtain a very good agreement as well for $\delta=3$ and $4$, when there is caging.
In these cases we also plot the prediction that would be made
were caging neglected (colored dashed lines), which is off by an order of magnitude.
The corresponding value of $\phicmono$ is $0.6137$.
The model compares as well \cite{supmat} to other data from the literature
\cite{chang1994effect,gondret1997}.

%%%%%%%%%%%%%%%%%%%%%%%%%%%%%%%%%%%%%%%%%%%%%%%%%%
% Conclusion
%
% Effective volume and jamming fraction
Throughout the paper, we have considered the effective volume of the particles, 
and its average \phistar over the size distribution.
It follows from Eq.~(\ref{eq:visc}) that \phistar is very similar to the \phic 
of the viscosity laws, although its interpretation is different.
\phistar is a measure of the topology of the microstructure,
and as such it depends on the flow that produces it.
It is therefore not surprising to obtain different values of \phistar between
pinch-off and rheometer experiments;
what matters is the variation of \phistar within a given experiment.
Although jamming would of course affect its value,
it has a physical meaning far from the jamming transition
(our experiments include the dilute range $\phi \leq 10\%$):
the effective volume in which particles amplify the fluid deformations.

% Caging
The caging of small particles by large ones has an important effect
on the local strain rate and on the bulk viscosity,
especially when there are few small particles.
Caging occurs when small particles can be trapped in the interstices between the large ones,
that is when the size ratio is greater than $\hdi \simeq 2.41$.
Because of its basic hypotheses, our model is only valid at low shear rates for which the
microstructure is relatively isotropic and contacts between the repulsive particles are unlikely.
In the general case, one should take into account the more complex shape of the interstices,
due to shear breaking symmetry, as well as the kinetics of caging and freeing.

% No need for friction...
The good agreement between our model -- that does not consider any kind of 
deformation nor dissipation due to solid friction between particles -- 
and the various rheological data presented here suggests
that the work of friction forces is negligible in our system,
in agreement with contact force measurements~\cite{nguyen-le2023}.
However, this does not mean that friction is negligible, or that there are no contacts, 
but only that these act principally by reducing particle mobility,
hence hindering the flow and increasing viscous dissipation
~\cite{jamali2019alternative,wang2020hydrodynamic}.
% This only concerns the viscosity at low shear rates; 
The question remains open for high shear rates, typically in the case of shear-thickening.
One would need to experimentally measure the fluid deformation in a suspension
undergoing shear-thickening, maybe using dissolved polymers.

\begin{acknowledgments}
    We thank J.~Morris and E.~Guazzelli for interesting discussions at the early stage 
    of the project.
    This material is based upon work supported by the National Science Foundation
    under NSF Faculty Early Career Development (CAREER) Program Award CBET No. 1944844.
    Contributions: VT and AS designed the study and wrote the manuscript;
    NV performed the pinch-off experiments;
    VT performed the viscosity measurements and developped the model.
\end{acknowledgments}

\bibliography{bidisperse}

\end{document}